\documentclass[12pt]{article}
\usepackage{graphicx}
\usepackage[cp1251]{inputenc}

\begin{document}
 \noindent {\footnotesize\it Astrophysical Bulletin, 2019, Vol. 74, No 1}
 \newcommand{\dif}{\textrm{d}}

 \noindent
 \begin{tabular}{llllllllllllllllllllllllllllllllllllllllllllll}
 & & & & & & & & & & & & & & & & & & & & & & & & & & & & & & & & & & & & & \\\hline\hline
 \end{tabular}

  \vskip 0.5cm
  \centerline{\bf\large Search for possibly evolutionary linked globular and open clusters}
  \bigskip
  \bigskip
  \centerline
 {V.V. Bobylev\footnote [1]{e-mail: vbobylev@gaoran.ru}
       and
  A.T. Bajkova}
  \bigskip

  \centerline{\small\it Main (Pulkovo) Astronomical Observatory, Russian Academy of Sciences,}

  \centerline{\small\it St.-Petersburg, 196140 Russia}
 \bigskip
 \bigskip
 \bigskip

 {
{\bf Abstract}---Based on a large sample of 133 Galactic globular
clusters we obtained a new estimate of the frequency of
globular-cluster impacts onto the Galactic plane, which we found
to be equal to three events per 1 Myr. Our computations involving
new kinematical data do not support the well-known hypothesis
about the possible origin of the open cluster Stephenson~2 as a
result of the massive globular cluster $\omega$ Cen crossing the
Galactic disk. Our results also do not support the well-known
hypothesis that the globular cluster NGC6397 could trigger the
formation of the open cluster NGC6231. We found for the first time
six globular clusters, which could have triggered the formation of
an open cluster when crossing the Galactic plane. These are the
globular clusters NGC104, NGC2808, NGC6362, NGC6540, NGC6749, and
NGC6752. For each of these clusters we identify one or several
open clusters, which were possibly born via such scenario. In our
opinion, of greatest interest are the pairs NGC104--Ruprecht 129,
and NGC6362--Pismis 11.
  }


 \subsection*{INTRODUCTION}
The crossing of the Galactic disk by a globular cluster (GC) can
stimulate star formation processes. This process may involve (a)
gravitational focusing where as a result of the approach of the GC
to the disk matter is pulled to a certain point and/or (b) strong
contraction of the disk matter in a certain direction. In the
latter case the strongest contraction effect is observed in the
case of the oblique incidence of a massive body, as established by
Comeron and Torra [1, 2] who modeled an impact of a massive
high-velocity cloud onto the Galactic disk. Furthermore, the
computations performed by Levy [3] showed that the passage of GC
through the Galactic disk may produce a shock wave. Wallin et al.
[4] showed, based on the model of the gravitational focusing of
the gas, that large OB-associations may form after about 30 Myr
after the GC passage through the Galactic disk.

Brosche et al. [5] appear to be the first to raise the issue about
finding some observational consequences of the Galactic-disk
crossings of two GCs, NGC362 and NGC6218. For example, the cluster
NGC362 should have crossed the disk at rather high velocity for
mass ejection from the disk to be possible.

To consider another scenario, Rees and Cudworth [6] proposed to
study the NGC6397--NGC6231 pair (a GC and a young open cluster).
The above authors associated the Galactic--disk crossing by the GC
with the subsequent formation of the open star cluster (OSC) at
the crossing place.

Vande Putte and Cropper [7] analyzed the Galactic trajectories of
54 GCs with measured proper motions, radial velocities, and
distances and found the NGC3201, NGC6397, and NGC6838 clusters to
be the promising sites for searching. Near the crossing places of
these GCs OB-associations are located, which could have formed as
a result of the impact of a GC onto the disk. Salerno et al. [8]
showed that the passage of the GC $\omega$ Cen through the
Galactic disk could have triggered the formation of the OSCs
Stephenson~2 and [BDS2003] 122.

GCs may be involved not only in the formation of OSCs, but also in
their destruction. De la Fuente Marcos et al. [9] computed the
parameters of mutual encounters between known Milky Way globular
and open clusters. The record breaker turned out to be the pair
consisting of the GC FSR 1767 (2MASS--GC04) and young OSC Ruprecht
127: their separation 22 million years ago was only 36 pc. The
above authors consider this pair as an example of the destruction
of clusters because they come too close to each other.

The results of such studies are very much dependent on the
measurement errors of the observed cluster parameters, in
particular, on the errors of their proper motions. Therefore the
publication of high precision proper motions for a number of
Milky-Way GCs based on Hubble Space Telescope observations
[10--12] served as a stimulus for carrying out this work.

The aim of this study is

• To estimate the frequency of the passages of GCs across the
Galactic midplane by numerically integrating orbits in the most
realistic model of the Galactic potential taking into account the
contribution of both axisymmetric components (bulge, disk, and
halo) and that of the bar and spiral structure with the parameters
determined from the most recent observational data;

• Validate some of the well-known results of various authors;

• And, most importantly, use modern data to search for other
possible cases of the formation of OSCs in the Galactic disk
triggered by the passage of a GC through it.

 \section*{THE DATA}
Our cluster sample is based on the MWSC (Milky Way Star Clusters)
catalog presented in [13]. It contains the positions, proper
motions, and radial velocities for stars both in open and globular
clusters of our Galaxy. When analyzing the positions of OSCs we
use more extensive catalog [14]. We usually excluded objects
marked by labels “a”, “n”, and “s” in these catalogs
(associations, nebulae, and asterisms).

The average proper motions of clusters in the MWSC are based on
the stellar proper motions of PPMXL catalog [15], which have large
random errors ranging from 4 to 10 mas yr$^{-1}$. One of the aims
of this paper is to construct the Galactic orbits of globular and
open clusters over long time intervals. This requires maximum
possible accurate knowledge of the initial kinematic parameters.
For this reason we adopted more reliable data for a number of GCs.

1. We used the data from [11], where the mean absolute proper
motions of ten selected Galactic bulge GCs are derived from both
ground-based observations made with European Southern Observatory
telescopes and from the data obtained with the Hubble Space
Telescope (HST). The average epoch difference was 25 years and the
average error of the inferred proper motion of a single cluster is
of about 0.5 mas yr$^{-1}$. These are GCs Terzan~1, Terzan~2,
Terzan~4, Terzan~9, NGC6522, NGC6540, NGC6558, NGC6652, NGC6681,
and Palomar~6. A detailed analysis of their individual orbits can
be found in [16]. For clusters Terzan~2, Terzan 4, NGC6652, and
Palomar~6 we applied corrections in accordance with [17].

2. We adopted the absolute proper-motion components and
heliocentric distance to $\omega$~Cen cluster inferred from HST
observations and reported in [12]:
\begin{equation}
 \begin{array}{rr}
 \mu_\alpha\cos\delta=-3.238\pm0.028~{\hbox {\rm mas yr$^{-1}$}},\\
           \mu_\delta=-6.716\pm0.043~{\hbox {\rm mas yr$^{-1}$}},\\
                    r= 5.20\pm0.25~{\hbox {\rm kpc}}.
 \label{Libralato-mu}
 \end{array}
 \end{equation}
The proper-motion components determined in (1) differ strongly
from those used in [8]
\begin{equation}
 \begin{array}{rr}
 \mu_\alpha\cos\delta=-5.08\pm0.35~{\hbox {\rm mas yr$^{-1}$}},\\
           \mu_\delta=-3.57\pm0.34~{\hbox {\rm mas yr$^{-1}$}},\\
                    r= 5.3\pm0.5~{\hbox {\rm kpc}}
 \label{Salerno-mu}
 \end{array}
 \end{equation}
when modeling the passage of the $\omega$ Cen cluster through the
Galactic disk. Therefore repeating the study carried out in [8],
which we perform in this paper using new data and a different
model of Galactic gravitational potential is of great interest.

3.We use the proper motions of theGCsNGC104, NGC5272, NGC6121,
NGC6397, and NGC6656 based on Gaia TGAS (Tycho--Gaia Astrometric
Solution) catalog data by Watkins and van den Marel [18]. The
above authors also report the average trigonometric parallaxes for
the five clusters. However, it is their proper motions that are of
primary importance. The distance estimates are not too precise
because they are based on the few stars in each cluster.
Furthermore, distances greater than 2 kpc so far cannot be
reliably determined because of the parallax errors in the current
GaiaDR1 release.

4. To complete the picture, we point out our use of new
measurements of the components of the absolute proper motion of
the NGC2419 cluster:
 $(\mu_\alpha\cos\delta,\mu_\delta)=(-0.17\pm0.26;-0.49\pm0.17)$~mas yr$^{-1}$,
based on HST and GaiaDR1 data by Massari et al. [19]. These values
differ strongly from those reported in catalog [13].

5. NGC6397 is one the closest GCs to the Sun. As we pointed out in
item 3 of this section, we use for this cluster the new precise
proper motion values from [18]. Furthermore, Brown et al. [20]
recently found a new estimate for the trigonometric parallax of
this cluster, $\pi_{trig}=0.418\pm0.018$~mas, based on HST
observations. As a result, we can now test the hypothesis of Rees
and Cudworth [6] at a new level of accuracy.

 \section*{THE METHOD}
The method that we use to search OSCs whose formation was likely
triggered by the impact of GC onto the Galactic disk is based on
integration of the orbits of globular and open clusters in the
Galactic gravitational potential. Integration of GCs orbits allow
us to determine coordinates and times of the passage of the GC
through the Galactic disk. Integration of OSCs orbits over a
certain interval allows us to determine how close they were to the
reconstructed impact sites and likely time of the birth of the
OSC. The use of statistical Monte Carlo method makes it possible
to construct an entire domain $(X,Y)$ of globular-cluster
crossings with the Galactic disk and also the set of orbits of the
OSCs considered taking into account the errors in the kinematical
data of objects and parameters of the Galactic model. This makes
it possible to estimate the degree of overlap of globular-cluster
impact domains with open-cluster orbits and hence the probability
of the birth of an OSC. Note that in all numerical simulations
performed within the framework of this paper we used 1000
Monte-Carlo iterations to construct the confidence domain of the
points where the GC crosses the Galactic disk and 100 iterations
when constructing the ensemble of open-cluster orbits. The MWSC
catalog does not give the random errors of the inferred cluster
distances and we set them equal to 10\% (except otherwise stated)
for all clusters of the this catalog. Note that the confidence
intervals of the reconstructed sets of (1) points where the GC
intersect with the Galactic plane and (2) open-cluster orbits
correspond to the probability of 99.7\% (3$\sigma$).

We describe our adopted model of the Galaxy below. We consider the
Navarro–Frank–White model [23] that we refined based on modern
data [21, 22] and extended with terms taking into account the
effect of the central bar and spiral density wave to be most
realistic.

\subsection*{Model of Axisymmetric Potential of the Galaxy}
Model of the axisymmetric gravitational potential of the Galaxy
has the form of the sum of three components—central spherical
bulge $\Phi_b(r(R,Z))$, disk $\Phi_d(r(R,Z))$, and massive
spherical dark-matter halo $\Phi_h(r(R,Z))$:
 \begin{equation}
 \begin{array}{lll}
  \Phi(R,Z)=\Phi_b(r(R,Z))+\Phi_d(r(R,Z))+\Phi_h(r(R,Z)).
 \label{pot}
 \end{array}
 \end{equation}
Here we use cylindrical coordinate system ($R,\psi,Z$) with the
origin at the Galactic center. In the Cartesian coordinate system
$(X,Y,Z)$ with the origin at the Galactic center the distance to a
star (spherical radius) is $r^2=X^2+Y^2+Z^2=R^2+Z^2$. The $X$-axis
points from the Galactic center to the Sun; the $Y$-axis is
perpendicular to the $X$-axis and points in the direction of
Galactic rotation, and the $Z$-axis is perpendicular to the
Galactic plane $XY$ toward the North Galactic pole. The
gravitational potential is given in the units of 100 km$^2$
s$^{-2}$, distances are in kpc, and masses in the units of the
mass of the Galaxy, $M_{gal}=2.325\times 10^7 M_\odot$,
corresponding to Galactic constant $G=1.$

The potentials of the bulge $\Phi_b(r(R,Z))$ and disk
$\Phi_d(r(R,Z))$ have the form proposed by Miyamoto and Nagai
[24]:
 \begin{equation}
  \Phi_b(r)=-\frac{M_b}{(r^2+b_b^2)^{1/2}},
  \label{bulge}
 \end{equation}
 \begin{equation}
 \Phi_d(R,Z)=-\frac{M_d}{\Biggl[R^2+\Bigl(a_d+\sqrt{Z^2+b_d^2}\Bigr)^2\Biggr]^{1/2}},
 \label{disk}
\end{equation}
where $M_b,$ $M_d$ are the masses of the components and $bb, ad,
bd$ are the scale parameters in kpc. The halo component is adopted
in the form of the Navarro–Frank–White model [23]:
 \begin{equation}
  \Phi_h(r)=-\frac{M_h}{r} \ln {\Biggl(1+\frac{r}{a_h}\Biggr)}.
 \label{halo-III}
 \end{equation}
Table 1 lists the values of the parameters of Galactic potential
model (4)--(6) found in [21, 22] based on circular rotation
velocities of Galactic objects located at distances out to
$R\sim200$~kpc. Note that the corresponding Galactic rotation
curve was constructed adopting the solar Galactocentric distance
of $R_0=8.3$~kpc and LSR velocity relative to the Galactic center
$V_0=244$ km s$^{-1}.$

 {\begin{table}[t]                                    
 \caption[]
 {\small\baselineskip=1.0ex
 Parameters of the Galactic potential model
 $M_{gal}=2.325\times10^7 M_\odot$
  }
 \label{t:model-III}
 \begin{center}\begin{tabular}{|c|c|r|}\hline
 Parameters       &     Values        \\\hline
 $M_b$(M$_{gal}$) &    443$\pm27$     \\
 $M_d$(M$_{gal}$) &   2798$\pm84$     \\
 $M_h$(M$_{gal}$) &  12474$\pm3289$   \\
 $b_b$(kpc)       & 0.2672$\pm0.0090$ \\
 $a_d$(kpc)       &   4.40$\pm0.73$   \\
 $b_d$(kpc)       & 0.3084$\pm0.0050$ \\
 $a_h$(kpc)       &    7.7$\pm2.1$    \\\hline
 \end{tabular}\end{center}\end{table}}

 \subsection*{Taking into Account the Bar and Galactic Spiral Pattern}
In the case of the Galactic bar and spiral pattern the
corresponding terms describing the potential of these components
are added to the right-hand side of formula (3). We adopted the
bar potential in the form of the triaxial ellipsoid model in
accordance with [25]:
\begin{equation}
  \Phi_B = -\frac{M_B}{(q_B^2+X^2+[Y\cdot a_B/b_B]^2+[Z\cdot a_B/c_B]^2)^{1/2}},
\label{bar}
\end{equation}
where $M_B$ is the mass of the bar, which is equal to $43.1\times$
M$_{gal}$; $a_B, b_B,$ and $c_B$ are the three semiaxes of the bar
$(a_B/b_B=2.381, a_B/c_B=3.03);$ $q_B$ is the length of the bar;
$X=R\cos\vartheta$ and $Y=R\sin\vartheta$, where
$\vartheta=\theta-\Omega_B \cdot t-\theta_B$, $\theta$ is the
initial position angle of the object: $\tan \theta=Y_0/X_0$
($X_0,Y_0$ are the initial coordinates of the object in the
Cartesian coordinate system in accordance with (11)), $\Omega_{B}$
is the circular velocity of the bar, $t$ is time, $\theta_B$ is
the bar orientation angle relative to Galactic axes $X,Y$, which
is counted from the line connecting the Sun and the Galactic
center (the $X$-axis) to the major axis of the bar in the
direction of Galactic rotation. We adopted the estimates of the
bar parameters $\Omega_B=55$~km s$^{-1}$ kpc$^{-1}$, $q_B=8$~kpc,
and $\theta_B=45^{\circ}$ from [26].

The potential of spiral density wave [27, 28] is described by the
following formula [29]:
\begin{equation}
 \Phi_{sp} (R,\theta,t)= A\cos[m(\Omega_p t-\theta)+\chi(R)],
 \label{Potent-spir}
\end{equation}
where
 $$
 A= \frac{(R_0\Omega_0)^2 f_{r0} \tan i}{m},
 \qquad
 \chi(R)=- \frac{m}{\tan i} \ln\biggl(\frac{R}{R_0}\biggr)+\chi_\odot,
 $$
Here $A$ is the amplitude of the spiral-wave potential; $f_{r0}$
is the ratio of the radial component of the perturbation from
spiral arms to the general gravitational attraction of the Galaxy;
$\Omega_p$ is the angular velocity of rigid rotation of the wave;
$m,$ the number of spiral arms; $i,$ the pitch angle of spiral
arms ($i<0$ for trailing arms); $\chi$ is the phase of the radial
wave and in that case $\chi=0^\circ$ corresponds to the arm
center; $\chi_\odot$ is the radial phase of the Sun with respect
to the spiral arm. We adopted the following values for the
parameters of the spiral wave:
 \begin{equation}
 \begin{array}{lll}
 m=4,\\
 i=-13^\circ,\\
 f_{r0}=0.05,\\
 \chi_\odot=-120^\circ,\\
 \Omega_p=20~\hbox {km s$^{-1}$ kpc$^{-1}$}.
 \label{param-spiral}
 \end{array}
 \end{equation}

 \subsection*{Equations of Motion}
Equations of motion of a test particle in the Galactic potential
have the following form:
\begin{equation}
 \begin{array}{llllll}
 \dot{X}=p_X,\quad
 \dot{Y}=p_Y,\quad
 \dot{Z}=p_Z,\\
 \dot{p}_X=-\partial\Phi/\partial X,\\
 \dot{p}_Y=-\partial\Phi/\partial Y,\\
 \dot{p}_Z=-\partial\Phi/\partial Z,
 \label{eq-motion}
 \end{array}
\end{equation}
where $p_X, p_Y,$ and $p_Z$ are the canonical moments and dot
denotes differentiation with respect to time. We integrate
equations (10) using the fourth-order Runge–Kutta algorithms.

We set the peculiar velocity of the Sun relative to the LSR equal
to $(u_\odot,v_\odot,w_\odot)=(11.1,12.2,7.3)$~km s$^{-1}$, in
accordance with [30]. Here we give the heliocentric velocities
measured in a moving Cartesian coordinate system: $u$ points
toward the Galactic center $v$ points toward Galactic rotation,
and $w$ is perpendicular to the Galactic plane in the direction
toward North Galactic pole.

Let us now denote the initial positions and spatial velocities of
test particles in the heliocentric coordinate system as
$(x_o,y_o,z_o,u_o,v_o,w_o)$. Then in the Galactic Cartesian
coordinate system the initial values of the positions and
velocities of test particles are given by the following formulas:
\begin{equation}
 \begin{array}{llllll}
 X_0=R_0-x_o, \quad Y_0=y_o, \quad Z_0=z_o+h_\odot,\\
 U=-(u_o+u_\odot),\\
 V=v_o+v_\odot+V_0,\\
 W=w_o+w_\odot,
 \label{init}
 \end{array}
\end{equation}
where $h_\odot=16$ pc is the height of the Sun above the Galactic
midplane according to the estimated given in [31].

 \section*{RESULTS AND DISCUSSION}
 \subsection*{Frequency of Globular-Cluster Impacts onto the Galactic Plane}
The MWSC provides the complete set of kinematic data for 133 GCs.
This set includes the estimate of the distance to the GC, its sky
coordinates, proper-motion components, radial velocity, and the
measurement errors of these quantities. We used these and refined
data (see “THE DATA” section) to integrate the Galactic orbit for
each GC 1 billion years into the past.

To estimate the frequency of GC impacts onto the Galactic plane,
we determine for each cluster the time instants when $Z=0$~kpc.
There are GCs, which have never reached the Galactic plane during
the last one billion years. Most of the GCs have crossed the
Galactic disk several times. Thus, e.g., $\omega$~Cen crossed the
Galactic disk 26 times during one billion years. The record
breaker is the GC VDBH229 (ESO 455-11), which is very close to the
Galactic center and which crossed the Galactic disk 175 times
during this time.

Fig. 1 shows the number of Galactic-plane crossings by the GCs of
our sample as a function of the distance to the rotation axis of
the Galaxy. The histograms shown in this figure are based on the
data of two types: (a) the data adopted from the MWSC catalog and
(b) the data refined as described in “THE DATA” section (the solid
line). As is evident from the figure, despite their scarcity, the
use of refined data affected the domain in the vicinity of the
Galactic center ($R<1$~kpc).

\begin{figure}[t]
{\begin{center}
   \includegraphics[width=0.5\textwidth]{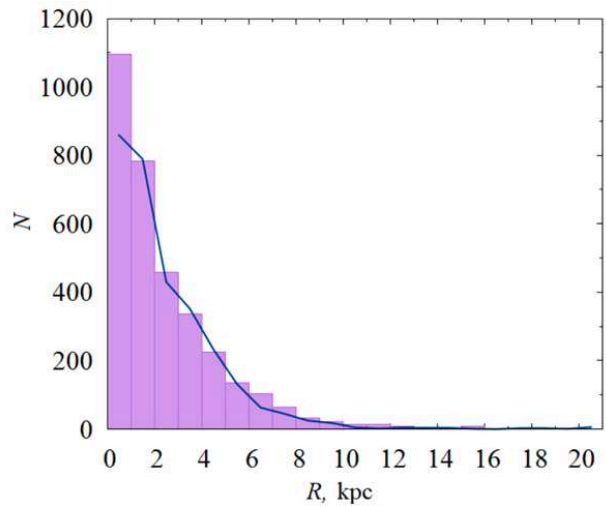}
 \caption{\small
Histogram (the filled bars) of the number of the crossings of the
Galactic plane $XY$ by GCs during the last billion years based on
the data of MWSC catalog [13]. The solid line shows the
distribution based on refined data. The horizontal axis gives
distance $R$ from the rotation axis of the Galaxy.
  } \label{f1}
\end{center}}
\end{figure}

A number of authors [7, 8] report, albeit without specifying the
method employed, an estimate of the frequency of GC impacts onto
the Galactic plane, which is of about one event in million
years.We found that, on the average, three crossings occur in one
million years within the entire disk.

When discussing the efficiency of the star-formation processes
associated with the effect of GCs onto the Galactic disk, Salerno
et al. [8] point out that effects of a GC impact onto the
formation of OSCs in the local disk region (within 1--2 kpc) is
highly likely in any place in the inner Galaxy about once every 10
million years. I.e., the above authors note the uniform nature of
star formation in the disk region. We, however, may suggest, based
on our inferred exponential decrease of the number of GC impacts
onto the disk shown in Fig.~1, that the efficiency of star
formation initiated by globular-cluster crossings of the Galactic
disk also deceases exponentially with the distance from the
Galactic center.

 \subsection*{Testing Two Well-Known Hypotheses}
Note that it takes considerable time after the GC impact onto the
Galactic plane for an OSC to form. Following [7], we use the
formula:
 \begin{equation}
 t=t_{\rm C}+t_{\rm SF}+t_{\rm A},
 \label{t-3}
 \end{equation}
where $t$ is the time elapsed since the GC crossed the Galactic
disk and until now; $t_C,$ the time interval since the crossing
time until the start of star formation; $t_{SF},$ the duration of
star formation, and $t_A,$ the age of the GC.

The value of the first term in formula (12) is known with large
uncertainty and lies in the interval from 0 to 30 Myr. For
example, according to the estimate reported in [32] and obtained
by modeling the impact of a high-velocity cloud onto the disk,
$t_C=15$~Myr, whereas according to [4], this time interval is
equal to $t_C=30$~Myr. The second term is, according to [33],
equal to $t_{\rm SF}=0.2$~Myr (for a star with the mass of
$M>M_\odot$), which is small compared to other terms and hence
this term can be neglected in the first coarse estimates.

We performed Monte-Carlo statistical simulations for each object
studied (GC or OSC). We added normally distributed random errors
with zero means and known standard deviations to the coordinates
of the object $(X,Y,Z),$ its spatial velocities $(U,V,W),$ and
parameters of the Gravitational potential model (see Table 1). We
determined the velocities, coordinates, and their errors by
applying Monte-Carlo statistical modelling method to measured
heliocentric distances, proper motions, and radial velocities of
the objects.

 \subsubsection*{The $\omega$~Cen--Stephenson 2 Pair}
Fig. 2 shows two diagrams obtained with two sets of measurement
data for the GC $\omega$~Cen and OSC Stephenson 2, which we
describe below. In these plots we show the (1) confidence
intervals of the domain of points where the GC $\omega$~Cen
crossed the Galactic plane $XY$ (the gray circles) obtained using
Monte-Carlo method taking into account the measurement errors of
the GC parameters and (2) the orbits of the OSC Stephenson 2
computed over the time interval from the present time to time $t$
of the GC crossing of the Galactic disk, also obtained using
Monte-Carlo method. These plots give the idea of the degree of
overlap between the set of points of the GC crossings of the
Galactic disk and the set of end points of open-cluster
trajectories at time $t.$ If such sets overlap then we can
conclude that the formation of the OSC could be triggered by the
GC. In the absence of such overlap the conclusion is negative.
Such an approach was implemented in the study of the other GC--OSC
pairs described below.

We already discussed the parameter values for the GC $\omega$~Cen
above in “THE DATA” section. Here we consider the parameters of
the OSC Stephenson 2. According to estimates [34], the age of
Stephenson 2 is equal to $t_{\rm A}\sim20$~Myr and it is located
at heliocentric distance of $d=5.9$~kpc. They are also consistent
with more recent estimates [35]: the age is in the $t_A=12-17$~Myr
interval and the distance is $d=5.8^{+1.9}_{-0.8}$~kpc. It is
important that Davies et al. [35] measured radial velocities for
26 stars of the OSC Stephenson 2 and determined the average radial
velocity of the OSC $V_{\rm LSR}\approx110\pm4$~km s$^{-1}$, which
we used together with the heliocentric distance $d=5.8\pm0.6$~kpc
for analyzing the 3D motion of this cluster. We combined these
with two sets of proper motion values for Stephenson 2:
\begin{equation}
 \begin{array}{rr}
 \mu_\alpha\cos\delta= 2.69\pm1.43~{\hbox {\rm mas yr$^{-1}$}},\\
           \mu_\delta=-7.00\pm1.43~{\hbox {\rm mas yr$^{-1}$}}
 \label{Kharch-mu}
 \end{array}
 \end{equation}
according to [13] and
\begin{equation}
 \begin{array}{rr}
 \mu_\alpha\cos\delta=-0.44\pm1.21~{\hbox {\rm mas yr$^{-1}$}},\\
           \mu_\delta= 0.80\pm0.53~{\hbox {\rm mas yr$^{-1}$}}
 \label{Dias-mu}
 \end{array}
 \end{equation}
according to catalog [36]. In catalog [36] the average proper
motion of the OSC are based on the proper motions of the UCAC4
catalog [37] with errors ranging from 1 to 4 mas yr$^{-1}$, which
are substantially smaller than the corresponding errors in the
PPMXL catalog [15].

\begin{figure}[t]
{\begin{center}
   \includegraphics[width=0.99\textwidth]{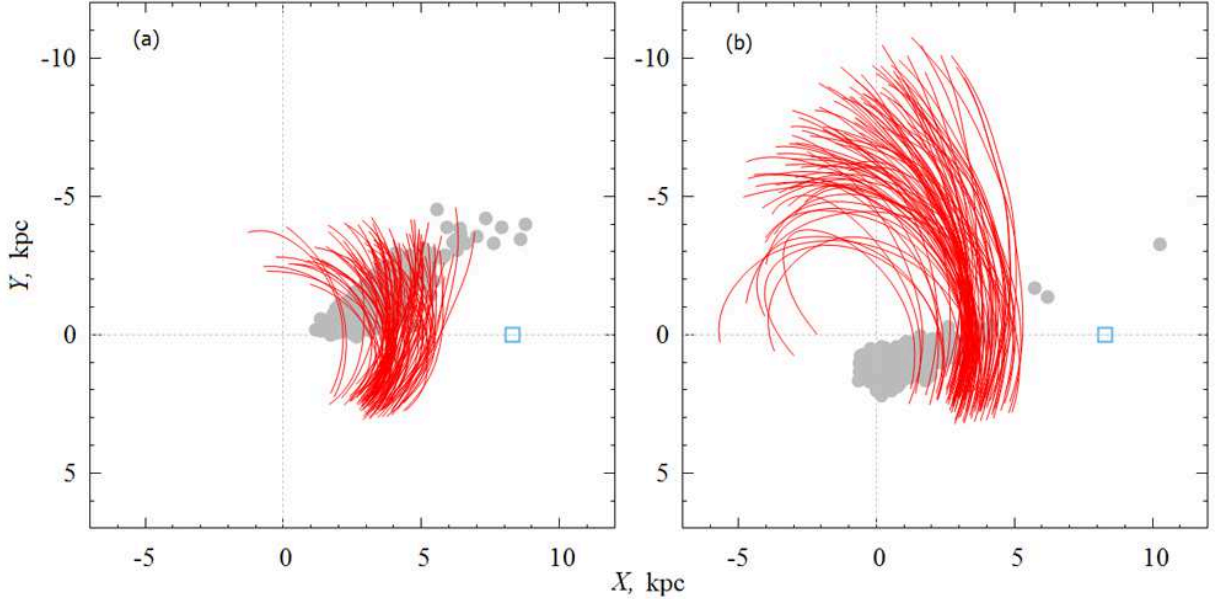}
 \caption{\small
Confidence intervals of the intersection of $\omega$~Cen GC with
the Galactic plane $XY$ (the gray circles) and ensembles of orbits
of the OSC Stephenson 2, (the solid lines) obtained using
Monte-Carlo method over the time interval from the last GC
crossing of the Galactic disk to the present time based on “old”
(a) and “new” (b) data. The Galactic center is located at the
origin and the square indicates the position of the Sun.
Currently, the Stephenson 2 cluster is located in the first
Galactic quadrant.
  } \label{f2}
\end{center}}
\end{figure}

As we already pointed out above, Fig. 2 shows the results obtained
for two sets of data used to analyze the $\omega$~Cen--Stephenson
2 pair. We refer to the data combination (2) for $\omega$~Cen and
(13) for Stephenson 2 as the “old” data and the combination (1)
and (14) as the “new” data.

Note that our approach differs from that of Salerno et al. [8] in
that we computed the spatial velocities of Stephenson 2 using real
measurements of both the proper motions and radial velocity. The
above authors [8] computed the model velocities of Stephenson 2
assuming that it moves in a purely circular orbit about the
Galactic center.

Based on the “old” data we found that $\omega$~Cen crossed the
Galactic plane 27.5 million years ago. This is the nominal value
and it is close to the $t=24$ Myr estimate obtained in [8]. Given
that the age of Stephenson 2 is $t_A=20$~Myr [34] we can conclude
that the hypothesis of Salerno et al. [8] that the formation of
this OSC may have been initiated by the impact of the GC
$\omega$~Cen onto the Galactic disk is true if it took the OSC 7.5
Myr to form in accordance with formula (12). Fig. 2a serves as a
weak confirmation of this hypothesis: the set of the end points of
the open-cluster trajectories corresponding to time $t=27.5$~Myr
overlaps, albeit only slightly, with the domain of the
globular-cluster crossings points with the Galactic disk. In this
case the nominal distance from the OSC orbit to the GC at the time
of the disk crossing was 2.49 kpc (see Table 2).

The situation is radically different if we use the “new” data set
for the GC $\omega$~Cen and OSC Stephenson 2. In this case the
nominal estimate of the time of crossing of the disk by the GC is
$t=45.5$~Myr and we integrate the orbits of the OSC backwards over
this time. As is evident from Fig.~2b, the end points of the
orbits of the OSC Stephenson 2 pass rather far from the domain of
intersection of the GC $\omega$~Cen with the $XY$ plane: the
nominal distance between the open-cluster orbit and the GC at the
time of the disk crossing is 9.59 kpc (see Table 2). We can
conclude from this that the hypothesis of Salerno et al. [8] is
strongly inconsistent with the new data about the GC $\omega$~Cen
and OSC Stephenson 2.

It is interesting that according to [13], the age of Stephenson 2
is $t_A=1$~Myr. By combining this age estimate with the “new” data
set we obtain too large time difference $t-t_A=44.5$~Myr, which is
far beyond our adopted interval. This fact further strengthens our
conclusion that pairing these two clusters is wrong.

\begin{figure}[t]
{\begin{center}
   \includegraphics[width=0.5\textwidth]{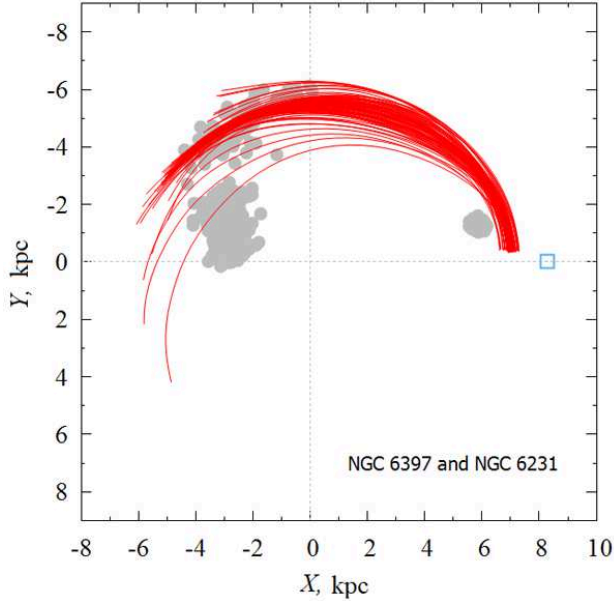}
 \caption{\small
Confidence domains of two successive crossings of the Galactic
plane $XY$ by the GC NGC6397 (the gray circles) and the ensemble
of orbits (the solid lines) of the OSC NGC6231 constructed over
the 49.5 Myr time interval backwards using Monte-Carlo method. In
the right-hand part of the figure we show the confidence domains
of the last crossing of the GC (3.5 million years ago), and in the
left-hand part, the confidence domains of the previous crossing
(49.5 million years ago). The Galactic center is at the origin and
the square indicates the position of the Sun. The OSC NGC6231 is
currently located in the second Galactic quadrant near the Sun.
  }
  \label{f3}
\end{center}}
\end{figure}

 \subsubsection*{The NGC6397--NGC6231 Pair}
According to the hypothesis of Rees and Cudworth [6], the passage
of the GC NGC6397 through the Galactic plane could trigger the
formation of the OSC NGC6231. The GC NGC6397 is also included into
list [7] as a candidate object for searching for any
manifestations of its Galactic-plane crossing. The above authors
analyzed the possibility of the formation of the OSC NGC6231 and
concluded that because of its young age ($t_A=3–5$~Myr) and quite
large distance from the site of the GC impact onto the disk it
could not form as a result of the passage of the GC NGC6397
through the Galactic plane, which, according to their estimate,
occurred 3.7 million years ago.

Our analysis of this pair led us to the same conclusion. The age
of the OSC NGC6231 is equal to $t_A=11.2$~Myr [13], and the last
time when the GC NGC6397 crossed the Galactic plane was about 3.5
million years ago. In this case condition (12) is evidently not
fulfilled.

Although the projections of the trajectories of the Galactic
orbits of the two clusters onto the $XY$ plane are close to each
other, they can be quite far away from each other in the 3D space.
Indeed, the vertical velocity $W=-4.3\pm0.5$ km s$^{-1}$ of the
OSC NGC6231 is small and therefore its orbit lies practically in
the Galactic plane. The corresponding velocity component of the GC
NGC6397 is much greater and equal to $W=-131.8\pm6.7$ km s$^{-1}$,
and therefore its orbit rises high above the Galactic plane (i.e.,
the GC may go far away from the OSC NGC6231 along the $Z$-axis).
Note the that random errors of the kinematical properties of both
clusters are small enough.

As we pointed out in “THE DATA” section, the proper motion of the
NGC6397 were computed by Watkins and van der Marel [18] based on
the data from Gaia TGAS catalog, i.e., they are sufficiently
reliable. The trigonometric parallax of the cluster was determined
by Brown et al. [20] with a fractional error of about 4\%. As a
result, statistical simulations produce compact confidence domains
for crossings of this GC.

After constructing the Galactic orbits of these two clusters over
the 50-Myr time interval backwards we found surprisingly that they
approach the GC at the time of the previous Galactic-plane
crossing 49.5 Myr ago to a nominal distance of about 3.55 kpc (see
Table 2, which gives the approach parameters both for the last
(case I) and the second last (case II) crossings). Fig.~3 shows
the confidence domains of the points of the crossing of the
Galactic plane $XY$ by the GC NGC6397 and the trajectory of the
OSC NGC6231 constructed over the 49.5 Myr interval backwards with
the allowance for data errors found via Monte-Carlo statistical
modeling method. In the right-hand part of the figure the gray
circles fill the domain of the last crossing of the Galactic disk
by the GC, which occurred 3.5 million years ago, and in the
left-hand part of the figure the gray circles show the domain of
the second-last crossing that occurred 49.5 million years ago. One
can see in the figure the overlap of the end points of the
open-cluster orbits and the domain of the previous Galactic-disk
crossing by the GC. This fact indicates that the effect of the GC
NGC6397 could, in principle, have triggered the formation of the
OSC NGC6231 about 49.5 million years ago during the previous GC
crossing of the Galactic plane. However, in this case the time
interval from the crossing to the start of star formation,
$t-t_A=38$~Myr, should have been unrealistically long. This fact
leads us to conclude that the NGC6397(II)--NGC6231 scenario is
also highly unlikely.

 \begin{table}[p]
 \caption[]
  {\small
 Parameters of the candidate GC--OSC pairs for past encounters that we found in this study
  }
  \begin{center}  \label{t:02}
  \small
  \begin{tabular}{|l|l|c|c|c|c|c|}
   \hline
    GC    &      OSC      & $\Delta r_t,$   &   $t,$ Myr &  $t_A,$  & $t-t_A,$ \\
          &               &           kpc   & in the past &  Myr &  Myr \\\hline
 $\omega$~Cen (old data)  & Stephenson2 & 2.49 & $ 27.5$ &     20.0 &      7.5 \\
 $\omega$~Cen (new data)  & Stephenson2 & 9.59 & $ 45.5$ &     20.0 &    $25.0$\\
 NGC 6397 (I) &  NGC 6231 &         0.96 &   3.5 &  11.2 & $-7.7$ \\       
 NGC 6397 (II)&  NGC 6231 &         3.55 &  49.5 &  11.2 &  38.3  \\\hline 

 NGC 104 & Ruprecht 129   &         0.25 &  52.5 &  31.6 &  20.9\\
 NGC 104 & FSR 52         &         0.52 &  52.5 &  31.6 &  20.9\\ 
 NGC 104 & Ruprecht 127   &         0.96 &  52.5 &  35.5 &  17.0\\
 NGC 104 & NGC 6396       &         0.90 &  52.5 &  32.0 &  20.5\\ 
 NGC 104 & Trumpler 27    &         1.44 &  52.5 &  38.0 &  14.5\\

 NGC 2808 & Pismis 20     &         1.39 &  41.5 &  31.6 &   9.9\\
 NGC 2808 & IRAS 6567     &         2.50 &  41.5 &  22.4 &  19.1\\ 
 NGC 2808 & Haffner 16    &         2.16 &  41.5 &  20.0 &  21.5\\

 NGC 6362 & Pismis 11     &         0.55 &  21.5 &   1.0 &  20.5\\
 NGC 6540 & [DBS2003] 102 &         0.57 &  15.5 &   1.3 &  14.2\\

 NGC 6541 & [DBS2003] 161 &         0.63 &  26.5 &   1.0 &  25.5\\
 NGC 6541 & [DBS2003] 102 &         1.23 &  26.5 &   1.3 &  25.2\\

 NGC 6749 (II) & Trumpler 27   &     1.23 &  51.5 &  38.0 &  13.5\\
 NGC 6749 (II) & Ruprecht 127  &     1.41 &  51.5 &  35.5 &  16.0\\
 NGC 6749 (II) & Turner 3      &     1.77 &  51.5 &  28.8 &  23.3\\
 NGC 6749 (II) & NGC 6396      &     1.88 &  51.5 &  32.0 &  19.5\\
 NGC 6749 (II) & Ruprecht 129  &     1.96 &  51.5 &  31.6 &  19.9\\
 NGC 6749 (II) & FSR 52        &     2.26 &  51.5 &  31.6 &  19.9\\

 NGC 6752      & [DBS2003] 115 &     0.90 &  47.5 &  24.0 &  23.5\\\hline

 \end{tabular}\end{center}
 \end{table}
\begin{figure}[t]
{\begin{center}
   \includegraphics[width=0.9\textwidth]{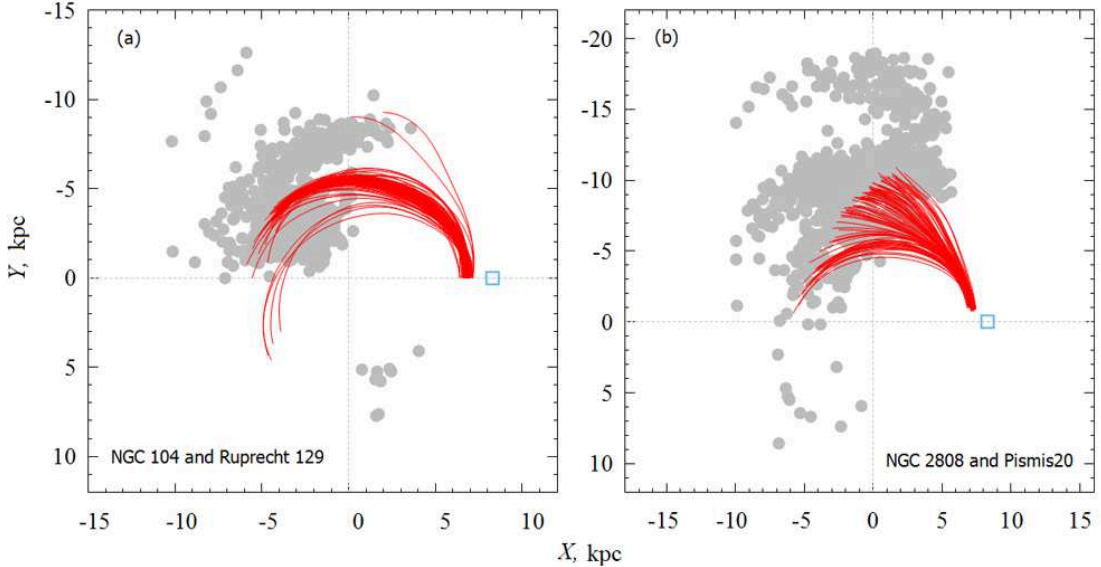} 
 \caption{\small
Confidence domains of the GC crossing points with the Galactic
plane $XY$ (the gray circles) and ensembles of open-cluster orbits
(the solid lines): (a) for the NGC104--Ruprecht 129 pair over the
52.5-Myr backward time interval obtained using Monte-Carlo method,
(b) for the NGC2808--Pismis 20 pair over the 41.5-Myr backward
time interval. The Galactic center is at the origin and the square
denotes the position of the Sun.
  }
  \label{f4}
\end{center}}
\end{figure}
\begin{figure}[t]
{\begin{center}
   \includegraphics[width=0.9\textwidth]{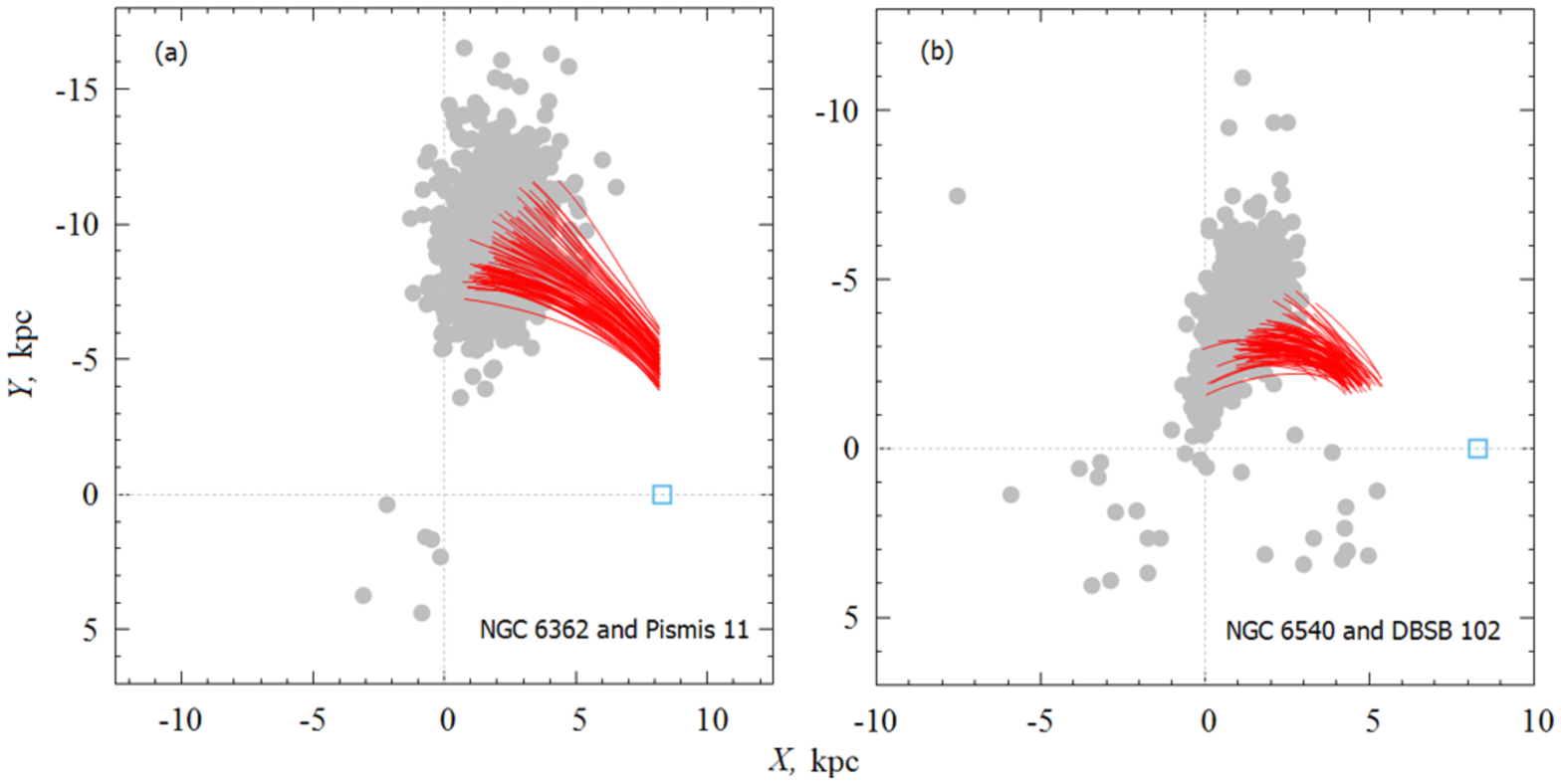} 
 \caption{\small
Confidence domains of the GC crossing points with the Galactic
plane $XY$ (the gray circles) and ensembles of open-cluster orbits
(the solid lines): (a) for the NGC6362--Pismis 11 pair over the
21.5-Myr backward time interval obtained using Monte-Carlo method,
(b) for the NGC6540--[DBS2003] 102 pair over the 15.5-Myr backward
time interval. The Galactic center is at the origin and the square
denotes the position of the Sun.
  }
  \label{f5}
\end{center}}
\end{figure}
\begin{figure}[t]
{\begin{center}
   \includegraphics[width=0.9\textwidth]{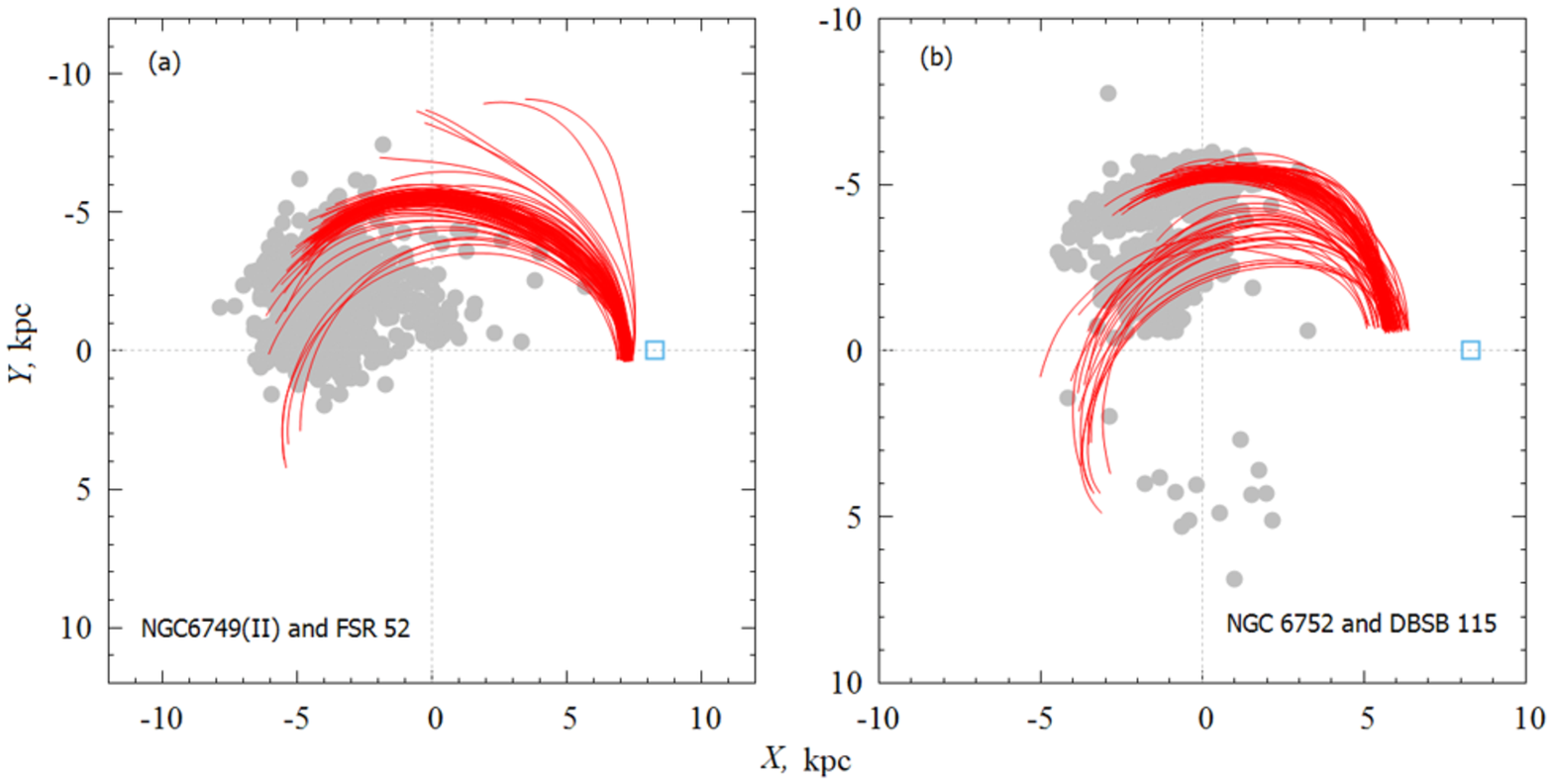} 
 \caption{\small
Confidence domains of the GC crossing points with the Galactic
plane $XY$ (the gray circles) and ensembles of open-cluster orbits
(the solid lines): (a) for theNGC6749(II)--FSR52 pair over the
51.5-Myr backward time interval obtained using Monte-Carlo method,
(b) for the NGC6752--[DBS2003] 115 pair over the 47.5-Myr backward
time interval. The Galactic center is at the origin and the square
denotes the position of the Sun.
  }
  \label{f6}
\end{center}}
\end{figure}

 \subsection*{Search for New Candidates}
We the searched for new candidate pairs of the form GC--OSC. To
this end, we composed, in addition to the already available sample
of 133 GCs, a sample of young $(\log t<7.6)$ OSCs with measured
proper motions, radial velocities, and distances. This sample
included 200 OSCs distributed over the entire Galactic disk. For
each pair we computed the so called encounter parameter $\Delta
r_t$, which is actually equal to the distance between the globular
and OSC at time $t,$ when the GC crossed the Galactic plane.

The results of the search obtained subject to the constraints
mentioned below are listed in Table 2. Note that the first four
rows of the table give the parameters of the two GC--OSC pairs
considered above. The second (bottom) part of the table presents
the new results.

We impose the following constraints: (1) $\Delta r_t$, should not
exceed 20\% of the heliocentric distance of the GC to the crossing
place (this constraint is based on the random error of photometric
distance), (2) the evident condition $t_A<t$ should be satisfied
in accordance with formula (12). It follows from Table 2 that even
with such conditions we have several open cluster candidates for
some GCs.

For the new candidate GCs we considered the OSCs associated with
the last crossing of the Galactic disk by the GC, and for the GC
NGC6749 we considered the OSCs associated with the place of the
second-last (II) crossing, which is more appropriate for the
formation of the OSC in terms of the temporal parameters.

Figs.~4--6 illustrate for a number of selected GC–OSC pairs the
degree of overlap between the confidence domains of the crossings
of the Galactic disk by the GC and the end points of the open
cluster orbits corresponding to the impact time. An analysis of
the modeling results obtained suggests that the birth of the OSC
could have been triggered by the GC. Below we give comments about
individual objects.

 \subsubsection*{NGC~104}
The GC NGC104 is located sufficiently close to the Sun
($d=4.50\pm0.05$~ kpc) and its kinematical properties are well
known. For this cluster we use the proper motion computed in [18]
based on Gaia TGAS catalog. The mass of NGC104 is
$0.84\times10^6~M_\odot$ [38], which is more than the masses of
most of the Galactic GCs with known mass estimates
$(1-3)\times10^5~M_\odot$ [7].

Fig. 4a shows the confidence domains of the NGC104
globular-cluster crossing points with the Galactic plane $XY$ and
trajectories of the OSC Ruprecht 129 computed over the 52.5-Myr
backward time interval. As is evident from the figure, we have
reliable overlap between the regions of the GC crossings of the
Galactic plane and the end points of the orbits of the OSCs
considered, providing a strong case for the hypothesis that the
formation of the OSC was triggered by the passage of the GC across
the Galactic disk.

Vande Putte and Cropper [7] determined for a number of GCs the
expected positions of OSCs at present assuming that their
formation was triggered by the last passage of the GC through the
disk. They give confidence intervals for each such position and,
as direct comparison of the positions has shown, all the OSCs
listed in Table 2 are either inside or close to the confidence
domain found in [7] for the GC NGC104. It is interesting that
Vande Putte and Cropper [7] used several models of the Galactic
gravitational potential. For example, in the Flynn potential model
(they denote this model as FL) they found for NGC104 the time of
the last Galactic plane crossing to be $t=52.6\pm1.9$~Myr, which
agrees well with our crossing time of $t=52.5$~ Myr.

The coordinates of all OSCs mentioned in Table 2 are close to each
other. However, the pair of OSCs with exactly the same
ages--Ruprecht 129 and FSR52--stands out and therefore. Therefore
it is appropriate to raise the question about their simultaneous
origin.

 \subsubsection*{NGC~6352 and NGC~6362}
Fig. 5a illustrates the case of the NGC6362--Pismis 11 pair. All
parameters appear to be acceptable for this pair: we are dealing
with a rather close encounter $\Delta r_t=0.55$~kpc and realistic
time interval $t-t_A=20.5$~Myr. As is evident from the figure, all
end points of open-cluster trajectories are located inside the
confidence domain of the GC crossing points, which, in addition,
is more compact compared to other cases. From this viewpoint the
NGC6362--Pismis 11 pair is one of the best candidates in our list.

Fig. 5b illustrates the case of the NGC6540--[DBS2003] 102 pair.
It stands out by its compact domain of crossing points and the
fact that practically 100\% of the end points of the open-cluster
trajectories are located inside this domain. The pair underwent a
rather close encounter, $\Delta r_t=0.57$~kpc, and is
characterized by rather short time interval $t-t_A=14.2$~Myr.

 \subsubsection*{NGC6749 and NGC6752}
A noteworthy feature of the GC NGC6749 is that almost all its
purportedly associated OSCs listed in Table 2 came close not only
to the place of the second-last crossing, but also to the place of
the last Galactic-disk crossing of the GC NGC104.

Fig. 6a illustrates the case of the NGC6749 (II)--FSR52, and
Fig.~6b, that of the NGC6752--[DBS2003]~115 pair. Both of these
pairs are characterized by low probability, of about 50\%, of
open-cluster trajectories getting into the domain of GC crossing
points.

 \section*{CONCLUSIONS}
We computed the Galactic orbits of 133 GCs by integrating their
trajectories in the Galactic gravitational potential that we
earlier refined based on modern data.We obtained a new estimate
for the frequency of globular-cluster impacts onto the Galactic
plane, which is equal to three events in 1 million years. We
showed that the number of such events is distributed exponentially
with a maximum in the central part of the Galaxy and rapidly
declines with the distance from the Galactic center. One should
therefore expect the efficiency of star formation initiated by
“bombardment” of the Galactic disk by GCs should also depend
strongly on Galactocentric distance. Another conclusion that
follows from it is that it is more efficient to search for
possible GC--OSC pairs in the inner regions of the Galaxy, and
such a search will be more resultative after the completion of the
Gaia space mission [39].

We tested two well-known hypotheses about the formation of an OSC
as a result of a Galactic-disk crossing by a GC.

The use of modern kinematical data did not support the hypothesis
of Salerno et al. [8] that the OSC Stephenson 2 could form after
the passage of the massive GC $\omega$~Cen through the Galactic
disk.

The hypothesis of Rees and Cudworth [6] that the passage of the GC
NGC6397 through the Galactic plane could have triggered the
formation of the OSC NGC6231 could be true not for the last
Galactic-disk crossing of NGC6397, which occurred 3.5 million
years ago as the authors of the hypothesis believed, but for the
previous passage, which occurred about 49.5 million years ago.
However, this would require unrealistically long time interval
between the crossing and the beginning of star formation,
$t-t_A=38$~Myr. This leads us to conclude that the scenario of
Rees and Cudworth [6] is unlikely.

We searched for new GC--OSC candidate pairs using the sample of
133 GCs and 200 OSCs and assessing the GC–OSC encounters by
analyzing each such pair. As a result, we identified six GCs
indicating for each one of them one or several OSCs potentially
paired with them (we list all these clusters in Table 2) within
the framework of the scenario where the formation of an OSC could
have been caused by the passage of a massive GC through the
Galactic disk. These GCs are NGC104, NGC2808, NGC6362, NGC6540,
NGC6749, and NGC6752. We found the pair with the closest distance
between the GC and OSC at the time of the Galactic disk crossing
by the GC to be NGC104--Ruprecht 129, which also has compatible
temporal parameters. Thus it is likely that the Galactic-disk
crossing of the GC NGC104 52.5 Myr ago could have served as the
mechanism that initiated star formation at the crossing site.
After 20.9 Myr the OSC Ruprecht 129 formed whose age is estimated
at 31.6 Myr. Also noteworthy is the GC--OSC pair
NGC6362--Pismis11, which has consistent parameters for the
realization of a similar scenario.

 \subsubsection*{ACKNOWLEDGMENTS}
We are grateful to the reviewer for useful comments, which helped
to improve the paper. This work was supported by Program P--28 of
the Presidium of the Russian Academy of Sciences, subprogram
“Space: the study of fundamental processes and their
interactions”.

 \bigskip
 \bigskip\medskip{\bf REFERENCES}
{\small

 1. F. Comeron and J. Torra, Astron. and Astrophys. 261, 94 (1992).

 2. F. Comeron and J. Torra, Astron. and Astrophys. 281, 35 (1994).

 3. V. V. Levy, Astron. Astrophys. Transactions 18, 621 (2000).

 4. J. F. Wallin, J. L. Higdon, and L. Staveley-Smith, Astrophys. J. 459, 555 (1996).

 5. P. Brosche, H.-J. Tucholke, A. R. Klemola, et al., Astron. J. 102, 2022 (1991).

 6. R. F. Rees and K. M. Cudworth, BAAS 35, 1219 (2003).

 7. D. Vande Putte and M. Cropper, Monthly Notices Royal Astron. Soc. 392, 113 (2009).

 8. G. M. Salerno, E. Bica, C. Bonatto, and I. Rodrigues, Astron. and Astrophys. 498, 419 (2009).

 9. R. de la Fuente Marcos, C. de la Fuente Marcos, and D. Reilly, Astrophys. and Space Sci. 349, 379 (2014).

 10. D. Massari, A. Bellini, F. R. Ferraro, et al., Astrophys. J. 779, 81 (2013).

 11. L. J. Rossi, S. Ortolani, B. Barbuy, et al., Monthly Notices Royal Astron. Soc. 450, 3270 (2015).

 12. M. Libralato, A. Bellini, L. R. Bedin, et al., Astrophys. J. 854, 45 (2018).

 13. N. V. Kharchenko, A. E. Piskunov, E. Schilbach, et al., Astron. and Astrophys. 558, A53 (2013).

 14. N. V. Kharchenko, A. E. Piskunov, E. Schilbach, et al., Astron. and Astrophys. 585, A101 (2016).

 15. S. Roeser, M. Demleitner, and E. Schilbach, Astron. J. 139, 2440 (2010).

 16. V. V. Bobylev and A. T. Bajkova, Astronomy Reports 61, 551 (2017).

 17. A. P\'erez-Villegas, L. Rossi, S. Ortolani, et al., Publ. Astron. Soc. Australia 35, e021 (2018).

 18. L. L. Watkins and R. P. van der Marel, Astrophys. J. 839, 89 (2017).

 19. D. Massari, L. Posti, A. Helmi, et al., Astron. and Astrophys. 598, L9 (2017).

 20. T. M. Brown, S. Casertano, J. Strader, et al., Astrophys. J. 856, L6 (2018).

 21. A. T. Bajkova and V. V. Bobylev, Astronomy Letters 42, 567 (2016).

 22. A. Bajkova and V. Bobylev, Open Astronomy 26, 72 (2017).

 23. J. F. Navarro, C. S. Frenk, and S. D. M. White, Astrophys. J. 490, 493 (1997).

 24. M. Miyamoto and R. Nagai, Publ. Astron. Soc. Japan 27, 533 (1975).

 25. J. Palous, B. Jungwiert, and J. Kopecky, Astron. and Astrophys. 274, 189 (1993).

 26. V. V. Bobylev and A. T. Bajkova, Astronomy Letters 42, 228 (2016).

 27. C. C. Lin and F. H. Shu, Astrophys. J. 140, 646 (1964).

 28. C. C. Lin, C. Yuan, and F. H. Shu, Astrophys. J. 155, 721 (1969).

 29. D. Fern\'andez, F. Figueras, and J. Torra, Astron. and Astrophys. 480, 735 (2008).

 30. R. Sch\"o nrich, J. Binney, and W. Dehnen, Monthly Notices Royal Astron. Soc. 403, 1829 (2010).

 31. V. V. Bobylev and A. T. Bajkova, Astronomy Letters 42, 1 (2016).

 32. J. R. D. Lepine and G. Duvert, Astron. and Astrophys. 286, 60 (1994).

 33. C. F. McKee and J. C. Tan, Nature 416, 59 (2002).

 34. S. Ortolani, E. Bica, B. Barbuy, and Y. Momany, Astron. and Astrophys. 390, 931 (2002).

 35. B. Davies, D. F. Figer, R.-P. Kudritzki, et al., Astrophys. J. 671, 781 (2007).

 36. W. S. Dias, H. Monteiro, T. C. Caetano, et al., Astron. and Astrophys. 564, A79 (2014).

 37. N. Zacharias, C. T. Finch, T. M. Girard, et al., Astron. J. 145, 44 (2013).

 38. A. Bellini, P. Bianchini, A. L. Varri, et al., Astrophys. J. 844, 167 (2017).

 39. Gaia Collab., T. Prusti, J. H. J. de Bruijne, et al., Astron. and Astrophys. 595, A1 (2016).

 }

  \end{document}